\documentclass[preprints,article,accept,pdftex,moreauthors]{Definitions/mdpi} 

\firstpage{1} 
\pubvolume{1}
\issuenum{1}
\articlenumber{0}
\pubyear{2023}
\copyrightyear{2023}
\datereceived{ } 
\daterevised{ } 
\dateaccepted{ } 
\datepublished{ } 
\hreflink{https://doi.org/} 

\usepackage{braket}
\newcommand\numberthis{\addtocounter{equation}{1}\tag{\theequation}}

\Title{Unimodular approaches to the cosmological constant problem}

\TitleCitation{Unimodular approaches to the cosmological constant problem}

\Author{Pavel Jirou\v{s}ek $^{1}$\orcidA{}*}

\AuthorNames{Pavel Jirou\v{s}ek}

\AuthorCitation{Jirou\v{s}ek, P.}

\address[1]{%
$^{1}$ \quad \it High Energy Physics, Cosmology \& Astrophysics Theory Group, University of Cape Town,\\Private Bag, Cape Town, South Africa, 7700; pavel.jirousek@uct.ac.za}

\corres{}

\firstnote{} 
\secondnote{}

\abstract{We review selected aspects of unimodular gravity and we discuss its viability as a solution of the old cosmological constant problem. In unimodular gravity the cosmological constant is promoted to a global degree of freedom. We highlight the importance of correctly setting up its initial data in order to achieve a resolution of the cosmological constant problem on a semi-classical level. We review recent path integral analysis of quantum aspects of unimodular gravity to note that the semi-classical findings carry over to the quantum level as well. We point out that a resolution of the problem inherently relies on a global constraint on the space-time four-volume. This makes the theory closely related to the vacuum energy sequester, which operates in a similar way. We discuss possible avenues of extending unimodular gravity that preserve the resolution of the cosmological constant problem.}

\keyword{unimodular gravity; cosmological constant problem; modified gravity; Weyl invariance} 

\begin{document}

\section{Introduction}
Prior to the measurement of the accelerated nature of the expansion of the Universe \cite{SupernovaSearchTeam:1998fmf,SupernovaCosmologyProject:1998vns} the cosmology community predominantly expected that the effective value of the cosmological constant (CC) would be zero. However, even before we have been burdened with reconciling the puzzling minuscule value of the CC, that we observe today \cite{Planck:2018vyg}, it has been recognized that there is an underlying problem with the vanishing of the vacuum energy \cite{Zeldovich:1968ehl,Sakharov:1967pk,Weinberg:1988cp}. This problem stems from the observation that the energy of the vacuum state of quantum fields behaves exactly as an effective cosmological constant. In the semi-classical approximation, where gravity behaves classically, while matter fields are quantized, these vacuum energies appear to be able to drive an accelerated expansion of the universe. However, any attempt at estimating these contributions have produced values of such magnitude that their effect on cosmology would be impossible to miss. It is needless to say that such effects have not been observed and while the vacuum energy has ultimately been measured to be non-zero it is still orders and orders of magnitude smaller than any estimation obtained from quantum field theory. The question then arises: Why do we not observe these large vacuum energies or rather what mechanism causes them to cancel out or vanish. For more details see i.e. \cite{Akhmedov:2002ts,Nobbenhuis:2004wn,Nobbenhuis:2006yf,Clifton:2011jh,Martin:2012bt}. Due to its origin, the above problem is often referred to as the 'old' cosmological constant problem. Note that more recent questions regarding the value of the vacuum energy like the coincidence problem often assume that the old cosmological constant problem is somehow solved and usually do not address it in any way.

The old cosmological constant problem is commonly considered to be a fine tuning problem as one can carefully tune the bare cosmological constant of general relativity (GR) in such a way that it cancels the quantum contributions up to the tiny residual amount that we observe. However, this view is grossly oversimplified. As it has been pointed out \cite{Burgess:2013ara,Kaloper:2013zca,Kaloper:2014dqa}, the cosmological constant also receives large contributions from higher loop correction of matter fields, which do not diminish with higher loop orders. One is then forced to tune the cosmological constant at every step of loop expansion to a very high degree of precision, which entails an infinite amount of fine tunings. Each as bad as the previous one. This signals that the running of such \emph{renormalized} cosmological constant is ultrasensitive to the UV completion of the matter theory, which we have no knowledge of. This is of course a disaster since it would imply that our understanding of gravity at the lowest energies depends significantly on the microscopic physics of large energies. Hence, we are in need to protect the effective cosmological constant from such effects.

In this paper we are going to discuss a popular theory commonly used to address this problem - the \emph{unimodular gravity} (UG) \cite{Buchmuller:1988wx,Buchmuller:1988yn,Carballo-Rubio:2022ofy}. The origins of this theory go back to Einstein himself who used the unimodular condition $\sqrt{-g}=1$ as a partial gauge fixing of diffeomorphism invariance to simplify calculations in GR \cite{Einstein:1916vd}. Only later it has been realized that assuming such a gauge fixing prior to variation of the Einstein-Hilbert action yields a modification of GR, where the trace of the Einstein equations is directly subtracted\footnote{We use the reduced Planck units $8\pi G_N=1$ and signature convention $\left(+,-,-,-\right)$.} \cite{vanderBij:1981ym,zee1985high,Buchmuller:1988wx,Buchmuller:1988yn,Ellis:2010uc}
\begin{equation}
    G_{\mu\nu}-\frac{1}{4}Gg_{\mu\nu}=T_{\mu\nu}-\frac{1}{4}Tg_{\mu\nu}\ .\label{sec1: traceless einstein equation}
\end{equation}
Surprisingly, these equations turn out to be classically equivalent to those of GR; however, with an unspecified cosmological constant. The key property of equations \eqref{sec1: traceless einstein equation} is that they appear to be blind to any cosmological constant term. This has raised hopes that within UG the quantum corrections to the vacuum energy would fail to affect the space-time geometry. This would clearly solve the old cosmological constant problem. Alas, this claim is often merely stated or it is considered to be an obvious fact, without providing any detail or references on the argument. Furthermore, it has been recently argued that the quantum corrections to vacuum energy in fact do not decouple in UG and that the old CC problem remains present in UG as well \cite{Nobbenhuis:2004wn,Padilla:2014yea,Padilla:2015aaa}. As far as we are aware the discussion on this topic is not settled and a consensus has not been reached. It is one of the aims of this work to highlight the origin in this difference in views.

In \cref{section: classical} we will discuss several popular formulations of UG and the status of the old CC problem within them. We show, that the resolution of whether UG solves the old CC problem or not, hinges on the way we provide initial data, that determine the effective cosmological constant. In particular, in formulations that rely on the use of a Lagrange multiplier to fix the unimodular condition for metric determinant, i. e.
\begin{equation}
    \lambda(\sqrt{-g}-1)\ ,
\end{equation}
specifying the initial value for the Lagrange multiplier directly, spoils the decoupling mechanism. Conversely, in formulations where the same restriction is achieved via a composite structure of the minimally coupled metric, for example:
\begin{equation}
    g^{phys}_{\mu\nu}=\frac{g_{\mu\nu}}{\sqrt[4]{-g}}\ ,
\end{equation}
such initial conditions cannot be given and the old CC problem appears to be resolved. We first discuss the ambiguity in the initial data on the level of equations of motion, then we provide a discussion for transverse diffeomorphism invariant formulations in \cref{section: unimodular constraint}. We review the fully diffeomorphism invariant theories \`{a} la Henneaux and Teitelboim \cite{Henneaux:1989zc} in \cref{section: HT formulations} and its Weyl invariant extensions \cite{Jirousek:2018ago,Hammer:2020dqp} in \cref{section: Weyl invariant}. We further comment on the usefulness of these extensions for further study of unimodular gravity. Finally, we comment on the appearance of a global constraint on four-volume of space-time in \cref{section: degrees of freedom}.

The problem with specifying the initial value of the Lagrange multiplier carries over to the quantum regime. We review a partial path integration procedure of the unimodular degrees of freedom that are extra in comparison to GR, for the Henneaux and Teitelboim formulation \cite{Henneaux:1989zc}. Quantum aspects of UG have been studied using path integral techniques in multiple recent works, e.i. \cite{Smolin:2009ti,Bufalo:2015wda,Padilla:2014yea,Saltas:2014cta,Percacci:2017fsy,Buchmuller:2022msj}. The integration can either be carried out, while keeping the initial value of the Lagrange multiplier fixed as in \cite{Padilla:2014yea,Bufalo:2015wda,Buchmuller:2022msj} or by keeping it free. The former reduces to GR with a directly specified cosmological constant. Hence, hence it corresponds to an extension of UG, which preserves it equivalence with GR. Consequently, the old CC problem persists. The latter calculation results in expression, which corresponds to GR with a CC that is selected by a global constraint on the space-time four-volume. Such fixing is inherently invariant under quantum corrections to vacuum energy as has been pointed out in \cite{Kaloper:2014dqa}. Hence the old cosmological constant problem seems to be alleviated. We also briefly discuss the appearance of quantum fluctuations of the cosmological constant that naively appears due to the promotion of CC to a degree of freedom that have been noted in \cite{Jirousek:2020vhy}.

Lastly, we review the proposal of vacuum energy sequestering \cite{Kaloper:2013zca,Kaloper:2014dqa}, which also achieves the decoupling of the quantum corrections to CC. This mechanism relies on a similar blindness of the equations of motion to the vacuum energies as we find in UG. Unlike UG, this approach forcibly introduces a pair of global constraints, which completely removes the ambiguity in providing initial data in UG. Vacuum energy sequester can be formulated as a local theory \cite{Kaloper:2015jra} that is very similar to UG. We point out that the local approach again introduces the ambiguity in providing initial data, which affected the solution of the old cosmological constant problem. Finally, we discuss the relation between the local and global version and propose how such procedure can be applied in UG

\section{Classical formulations of Unimodular Gravity}\label{section: classical}
As we have alluded in the introduction the motivation for UG stems mainly from the observation that the trace-free Einstein equation \cite{vanderBij:1981ym,zee1985high,Buchmuller:1988wx,Buchmuller:1988yn,Ellis:2010uc}\footnote{A similar equation has been written down originally by Einstein himself \cite{Einstein:1919gv}, however, only for a priori trace-less energy momentum tensor (of radiation). Only later it has been realized that these equations describe UG.}
\begin{equation}
    G_{\mu\nu}-\frac{1}{4}Gg_{\mu\nu}=T_{\mu\nu}-\frac{1}{4}Tg_{\mu\nu}\ ,\label{sec2: traceless einstein equation}
\end{equation}
contains no information about the cosmological constant term in the Einstein-Hilbert action. Despite this, these equations are almost equivalent to standard Einstein's equations. We can see this by taking the covariant divergence of both sides, which gives
\begin{equation}
    \partial_{\mu}(G-T)=0\ .\label{sec2: Bianchi identity}
\end{equation}
This is a differential constraint, which can be easily solved as
\begin{equation}
    G-T=4\Lambda\ ,\label{sec2: cosmological integration constant}
\end{equation}
where $\Lambda$ is an integration constant. Plugging this into the original traceless equation \eqref{sec2: traceless einstein equation} yields the Einstein equations with a cosmological constant $\Lambda$
\begin{equation}
    G_{\mu\nu}=T_{\mu\nu}+\Lambda g_{\mu\nu}\ ,\label{sec2: einstein equation}
\end{equation}
A crucial difference in comparison to GR is that any value for $\Lambda$ is admissible here. In other words, any solution of Einstein equations with \emph{arbitrary cosmological constant} is a solution of the traceless equations \eqref{sec2: traceless einstein equation}. The key property responsible for this arbitrariness is that equations \eqref{sec2: traceless einstein equation} are invariant under constant shifts of vacuum energy
\begin{equation}
    T_{\mu\nu}\rightarrow T_{\mu\nu}+\rho_{vac}\ g_{\mu\nu}\ ,\label{sec2: vacuum shift}
\end{equation}
where $\rho_{vac}$ is a constant. Crucially, the shift in the energy-momentum tensor, which is generated by accounting for the quantum corrections to vacuum energy, has exactly the form \eqref{sec2: vacuum shift} and therefore the original equations \eqref{sec2: traceless einstein equation} are indeed blind to such corrections. The symmetry \eqref{sec2: vacuum shift} holds even in equation \eqref{sec2: Bianchi identity}, but it is finally broken once we specify the cosmological constant $\Lambda$ in \eqref{sec2: cosmological integration constant}. If we wish to evolve some initial conditions using trace-free equations \eqref{sec2: traceless einstein equation}, we would soon find that we need to specify the effective cosmological constant \eqref{sec2: cosmological integration constant} in order to get a unique solution. However, as it has been pointed out \cite{Padilla:2014yea,Padilla:2015aaa}, specifying this parameter immediately yields the equation \eqref{sec2: einstein equation}, which is just an Einstein equation with a cosmological constant. It was here, where the original cosmological constant problem arose in the first place. This seems to imply that we have not succeeded in solving the old CC problem, rather we have just shifted it one step away.

However, the situation is not completely hopeless as the way we have chosen $\Lambda$ is far from unique. If we first consider a splitting of the energy-momentum tensor into two pieces
\begin{equation}
    T_{\mu\nu}=\Tilde{T}_{\mu\nu}+T^{vac}_{\mu\nu}\ .
\end{equation}
where the second piece on the right hand side accounts for the constant vacuum energy
\begin{equation}
    T^{vac}_{\mu\nu}=\frac{1}{4}T^{vac}\,g_{\mu\nu}\ ,
\end{equation}
where $T^{vac}$ is a space-time constant. Furthermore, we assign any additional quantum corrections to the vacuum energy to $T^{vac}_{\mu\nu}$. Hence, $\Tilde{T}_{\mu\nu}$ receives no such contributions. Plugging this splitting into \eqref{sec2: traceless einstein equation} we will find that $T^{vac}_{\mu\nu}$ completely drops out of the equations and therefore we can write
\begin{equation}
    G_{\mu\nu}-\frac{1}{4}Gg_{\mu\nu}=\tilde{T}_{\mu\nu}-\frac{1}{4}\tilde{T}g_{\mu\nu}\ .\label{sec2: traceless einstein equation 2}
\end{equation}
Repeating the argument above we obtain a differential constraint
\begin{equation}
    \partial_{\mu}(G-\Tilde{T})=0\ ,\label{sec2: Bianchi identity 2}
\end{equation}
which we integrate as
\begin{equation}
    G-\tilde{T}=4\tilde{\Lambda}\ .\label{sec2: tilde cosmological integration constant}
\end{equation}
This yields an Einstein equation
\begin{equation}
    G_{\mu\nu}=\Tilde{T}_{\mu\nu}+\tilde{\Lambda} g_{\mu\nu}\ .\label{sec2: einstein equation tilde}
\end{equation}
We can see that now only $\Tilde{T}_{\mu\nu}$ appears in the above equation, and since $\Tilde{T}_{\mu\nu}$ does not, by definition, receive any quantum correction to the vacuum energy it follows that a choice of $\Tilde{\Lambda}$ is also stable. Hence, if we perform a measurement of the cosmological constant and interpret it as a parameter of this equation rather then \eqref{sec2: einstein equation}, we will obtain a constant stable under quantum corrections.

It seems that the cosmological constant problem in equations \eqref{sec2: traceless einstein equation} is not solved automatically but allows us a leeway in how we interpret the measurement of the cosmological constant. Some ways are stable, while others are not. This is in a stark contrast with GR where the cosmological constant is only interpreted as the bare coupling constant of the Einstein-Hilbert action and we do not have any choice in interpretation. Note that this does not imply that GR and UG are physically inequivalent on classical level. Rather, the change in description in UG allows us to interpret the measurement of cosmological constant in a different manner, where the old cosmological constant problem does not arise. Going forward we will see that this is the case in other formulations of unimodular gravity as well.

\subsection{The unimodular constraint}\label{section: unimodular constraint}
Maybe the most common formulation of UG in literature is based on the so called unimodular condition, from which UG gets its name
\begin{equation}
    \sqrt{-g}=1\ .\label{sec2.1: unimodular constraint}
\end{equation}
In order to actually modify the dynamics of GR this condition is enforced prior to variation of the Einstein-Hilbert (EH) action. This can be achieved in multiple ways; however, the most common one is to enforce it via a Lagrange multiplier directly in the action \cite{Buchmuller:1988wx,Buchmuller:1988yn}
\begin{equation}
    S[g,\lambda,\Psi]=\int_{\mathcal{M}} d^{4}x\left [\ -\frac{1}{2}\sqrt{-g}R+\lambda\left (\sqrt{-g}-1\right )\right ]+S_{matter}[g,\Psi]\ .\label{sec2.1: unimodular constraint action}
\end{equation}
Here $S_{matter}$ accounts for any matter field $\Psi$ that we consider along the gravitational sector and $\mathcal{M}$ is the spacetime region under consideration. A downside of this formulation is that the action clearly breaks the diffeomorphism invariance of the original action. This is because the metric density $\sqrt{-g}$ is set to be equal to a scalar quantity, in this case a unity. Hence the action is invariant only under transverse diffeomorphisms generated by $\xi^{\mu}$ satisfying
\begin{equation}
    \nabla_{\mu}\xi^{\mu}=0\ .
\end{equation}
Such diffeomorphisms indeed preserve the metric density since
\begin{equation}
    \delta_{\xi}\sqrt{-g}=\mathcal{L}_{\xi}\sqrt{-g}=\frac{1}{2}\sqrt{-g}\nabla_{\mu}\xi^{\mu}=0\ .
\end{equation}
Hence the symmetry group of this theory is substantially different from GR. Crucially, the only diffeomorphism breaking term depends only on $\lambda$, while the rest of the action is still diffeomorphism invariant. Consequently, all the matter and gravity equations of motion remain covariant. In particular, the Einstein equation implied by action \eqref{sec2.1: unimodular constraint action} is
\begin{equation}
    G_{\mu\nu}+\lambda\,g_{\mu\nu}=T_{\mu\nu}\ .\label{sec2.1: unimodular constraint eom}
\end{equation}
Clearly the Lagrange multiplier $\lambda$ plays the role of the cosmological 'constant', which, at this point, is a general scalar field. However, since the matter sector of the action is assumed to be diffeomorphism invariant, it follows that the right hand side is covariantly conserved. By taking the covariant divergence of both sides we find
\begin{equation}
    \partial_{\mu}\lambda=0\ .\label{sec2.1: constancy of cosmological constant}
\end{equation}
Therefore, consistency requires that $\lambda$ is indeed a constant. The unimodular constraint enforced by $\lambda$ can be viewed locally as a mere gauge choice. Hence, it naively seems that any solution of GR with the cosmological constant $\lambda$ in \emph{any} coordinates can be considered as a solution of the above unimodular equations. Indeed, any such solution can be locally transformed into coordinates such that \eqref{sec2.1: unimodular constraint} is satisfied.

Let us now discuss the fate of the quantum corrections to the vacuum energy in this formulation. As opposed to the previous trace-free equations \eqref{sec2: traceless einstein equation}, the cosmological constant in \eqref{sec2.1: unimodular constraint eom} is an independent field and enters the Einstein equations directly. Consequently, such variable appears to have a privileged status in the theory and it seems only natural to provide initial conditions for it in order to solve \eqref{sec2.1: constancy of cosmological constant}. This, however, directly leads to the old cosmological constant problem. To demonstrate this, we proceed with the only consistent initial condition. That is when $\lambda$ is a spatial constant $\Lambda$
\begin{equation}
    \lambda(t_{1})=\Lambda\ .\label{sec2.1: initial condition for lambda}
\end{equation}
The equation \eqref{sec2.1: constancy of cosmological constant} then immediately fixes $\lambda(t)=\Lambda$ for the rest of the time evolution. Consequently, we are left with standard GR with a cosmological constant $\Lambda$ and the old cosmological constant problem appears again. A common counter-argument to this conclusion is that any quantum correction $\rho_{vac}$ to the cosmological constant appears in the action coupled to $\sqrt{-g}$. Surely, we can decouple such terms from gravity by using the constraint to eliminate the $\sqrt{-g}$ dependence:
\begin{equation}
    \int_{\mathcal{M}} d^{4}x\left [\lambda\left (\sqrt{-g}-1\right )+\rho_{vac}\sqrt{-g}\right ]\rightarrow \int_{\mathcal{M}} d^{4}x\left [\lambda\left (\sqrt{-g}-1\right )+\rho_{vac}\right ]\ .\label{sec2.1: decoupling procedure}
\end{equation}
Doing so prior to the derivation of equations of motion eliminates any information about the quantum corrections in the Einstein equations! However, while this seems as a straightforward step, we must realize that using the constraint within the action necessarily entails a redefinition of the Lagrange multiplier $\lambda$. In this case this redefinition is the shift
\begin{equation}
    \lambda\rightarrow \lambda-\rho_{vac}\ .\label{sec2.1: multiplier shift}
\end{equation}
Consequently, any initial condition for $\lambda$ \eqref{sec2.1: initial condition for lambda} posed prior to the use of the constraint within the action is shifted exactly by the same amount $\rho_{vac}$ in the opposite direction
\begin{equation}
    \Lambda\rightarrow \Lambda +\rho_{vac}\ .\label{sec2.1: initial condition shift}
\end{equation}
Hence, the value of the cosmological constant $\Lambda$ clearly receives the quantum contributions. One could be tempted to argue that we should therefore set up the initial conditions for $\lambda$ only after we calculate the quantum corrections to vacuum energy and use the constraint to decouple them. However, this is no different from fine-tuning the cosmological constant at each level of the loop expansion, because we would need to set up the right initial value for each loop order separately. Hence this 'solution' amounts to an infinite amount of fine tunings.

Interestingly, we can use the constraint to modify the action in a more substantial way so that the quantum vacuum energy contributions decouple automatically. Indeed, consider the following substitution
\begin{equation}
    g_{\mu\nu}\rightarrow \hat{g}_{\mu\nu}=g_{\mu\nu}\,|g|^{-1/4}\ ,\label{sec2.1: g hat definition}
\end{equation}
which is carried out everywhere in the action outside of the constraint itself. This results in a theory with the following form
\begin{equation}
    S[g,\lambda,\Psi]=\int_{\mathcal{M}} d^{4}x\left [\ -\frac{1}{2}\hat{R}+\lambda\left (\sqrt{-g}-1\right )\right ]+S_{matter}[\hat{g},\Psi]\ .\label{sec2.1: unimodular constraint action renormalized decoupled}
\end{equation}
Here $\Tilde{R}$ is the scalar curvature evaluated using $\hat{g}_{\mu\nu}$. The variation of \eqref{sec2.1: unimodular constraint action renormalized decoupled} with respect to $g_{\mu\nu}$ yields the equations of motion
\begin{equation}
    \hat{G}_{\mu\nu}-\frac{1}{4}\hat{G}\hat{g}_{\mu\nu}+\lambda g_{\mu\nu}=\hat{T}_{\mu\nu}-\frac{1}{4}\hat{T}\hat{g}_{\mu\nu}\ ,\label{sec2.1: traceless einstein equation with cc}
\end{equation}
where the hat above $G_{\mu\nu}$ and $T_{\mu\nu}$ signifies that the tensors are evaluated using the metric $\hat{g}_{\mu\nu}$. However, upon the constraint \eqref{sec2.1: unimodular constraint} we have $\hat{g}_{\mu\nu}=g_{\mu\nu}$ and thus we can drop the hats in the above equation. Crucially, taking the trace of \eqref{sec2.1: traceless einstein equation with cc} immediately implies $\lambda=0$. We get this conclusion without ever specifying its initial conditions! In fact, specifying non-zero initial conditions for $\lambda$ is clearly inconsistent. Substituting $\lambda=0$ back into equation \eqref{sec2.1: traceless einstein equation with cc} yields the traceless Einstein equations. We see that $\lambda$ no longer plays the role of the cosmological constant. Furthermore, since the entire matter and gravitational Lagrangian depend strictly on $\hat{g}_{\mu\nu}$, any quantum correction will contribute as $\rho_{vac}\sqrt{-\hat{g}}$ to action \eqref{sec2.1: unimodular constraint action renormalized decoupled}. Since the novel composite metric $\hat{g}_{\mu\nu}$ has a unit determinant by construction
\begin{equation}
    \sqrt{-\hat{g}}=1\ ,\label{sec2.1: unimodularity of g hat}
\end{equation}
these contributions decouple trivially, without the need to use the constraint or, equivalently, redefine $\lambda$. Therefore, the energy momentum tensor that appears in trace-free equations is automatically \emph{free of any quantum corrections to vacuum energy}. A downside of this is that the trace free equations lack \emph{all} information about the cosmological constant and its effective value must be put in by hand as it is seemingly not tied to initial conditions of any fields.

Considering the action \eqref{sec2.1: unimodular constraint action renormalized decoupled}, it is clear that the constraint in \eqref{sec2.1: unimodular constraint action renormalized decoupled} is rendered unnecessary (at least on the classical level) thus we can remove it from the action to obtain, yet another formulation of UG
\begin{equation}
    S[g,\Psi]=-\frac{1}{2}\int_{\mathcal{M}} d^{4}x\,\hat{R}+S_{matter}[\hat{g},\Psi]\ ,\label{sec2.1: unimodular constraint action renormalized decoupled no constraint}
\end{equation}
Since all terms in the action now depend purely on $\hat{g}_{\mu\nu}$, the action has a manifest Weyl invariance under transformations of the metric
\begin{equation}
    g_{\mu\nu}\rightarrow\omega^{2}g_{\mu\nu}\ ,\label{sec2.1: weyl symmetry}
\end{equation}
where $\omega$ is an arbitrary non-zero function. Consequently, the resulting equations of motion associated with $g_{\mu\nu}$ are necessarily traceless. The equations of motion associated to this action are indeed the Einstein traceless equations \eqref{sec2: traceless einstein equation 2} taken together with the unimodularity condition \eqref{sec2.1: unimodular constraint}. Since the Lagrange multiplier is no longer present in this formulation, choosing its initial value is clearly impossible here.

It is interesting to note that the Weyl symmetry \eqref{sec2.1: weyl symmetry} arises only after we eliminate the 'cosmological constant' term in the action
\begin{equation}
    \lambda\,\sqrt{-g}\ .
\end{equation}
Hence the symmetry of the action is increased by having $\lambda=0$. This bears a striking resemblance to the technical naturalness \cite{tHooft:1979rat} for the cosmological constant; however, in this case the extra symmetry is a gauge symmetry rather then a regular symmetry and consequently the conclusion is not that $\Lambda$ is protected against quantum corrections but instead it is necessarily vanishing. It is important to note that the Weyl symmetry in UG does not become anomalous in quantum regime as it has been shown in \cite{PhysRevD.91.124071,Alvarez:2013fs}.

Note that the main difference between the original action \eqref{sec2.1: unimodular constraint action} and the theory \eqref{sec2.1: unimodular constraint action renormalized decoupled no constraint} is that in the former, the unimodular condition \eqref{sec2.1: unimodular constraint} is enforced via a Lagrange multiplier, while the latter achieves the same by universally coupling to a \emph{composite metric} \eqref{sec2.1: g hat definition}. It is not immediately clear why this should make a difference as a standard intuition dictates that these theories should be entirely equivalent. Yet as we have seen, they behave differently. The difference stems from the fact that in the theory \eqref{sec2.1: unimodular constraint action} we are tempted to introduce initial condition for the constant part of $\lambda$. This implies that that the zero mode is not varied and thus the integral conclusion of the unimodular condition
\begin{equation}
    \int_{\mathcal{M}}d^{4}x\sqrt{-g}=\int_{\mathcal{M}} d^{4}x\ ,\label{sec2.1: global constraint four volume}
\end{equation}
is not meant to be enforced. Note that this is the only diffeomorphism invariant information in \eqref{sec2.1: unimodular constraint} and hence it represents a physical constraint \cite{Percacci:2017fsy,deBrito:2021pmw}. It is thus not surprising that abandoning it leaves the theory unmodified - equivalent to GR. Leaving the initial value for $\lambda$ unspecified allows us to use the constraint \emph{freely} and hence we can use \eqref{sec2.1: decoupling procedure} without limits to decouple any contribution. Consequently, the cosmological constant problem is alleviated. The condition \eqref{sec2.1: global constraint four volume} should then provide the missing information in the trace-less Einstein equations \eqref{sec2: traceless einstein equation 2} and consequently allow us to to determine the effective cosmological constant. In the current setting this point is difficult to demonstrate; however, we will revisit it for the HT formulation in \cref{section: degrees of freedom}, where analogous situation occurs.

\subsection{Henneaux and Teitelboim UG}\label{section: HT formulations}
The introduction of the unimodularity condition \eqref{sec2.1: unimodular constraint} in the previous formulation has the unfortunate consequence of reducing the gauge group of the theory from diffeomorphisms to transverse diffeomorphisms. However, the full diffeomorphism invariance can be restored by introducing a novel vector density $V^{\mu}$. This construction has been described in \cite{Henneaux:1989zc} and the resulting theory is given by the following action\footnote{
This theory can be very easily rewritten in several other forms that are immediately equivalent. The only difference is that the fields $V^{\mu}$ and $\lambda$ can be redefined in such a way that the constraint part of the action becomes
\begin{equation}
    \sqrt{-g}\lambda(\nabla_{\mu}W^{\mu}-1)\ ,
\end{equation}
where $W^{\mu}$ is an ordinary vector field. Other popular choice is
\begin{equation}
    \lambda(\frac{1}{4!}\epsilon^{\mu\nu\sigma\rho}F_{\mu\nu\sigma\rho}-\sqrt{-g})\ ,
\end{equation}
where $F_{\mu\nu\sigma\rho}$ is the field strength of a 3-form gauge field $A_{\mu\nu\sigma}$ given as $F_{\mu\nu\sigma\rho}=4\partial_{[\mu}A_{\nu\sigma\rho]}$.
}, which is usually referred to as the Henneaux and Teitelboim (HT) unimodular gravity
\begin{equation}
    S_{HT}[g,\lambda,V]=\int_{\mathcal{M}} d^{4}x\Big [-\frac{1}{2}\sqrt{-g}R-\lambda\big ( \partial_{\mu}V^{\mu}-\sqrt{-g}\big )\Big ]+S_{matter}[g,\Psi]\ .\label{sec2.2: unimodular HT action}
\end{equation}
Note that the divergence of a vector density is a scalar density and therefore the above action is fully diffeomorphism invariant. We can clearly see that this action reduces to \eqref{sec2.1: unimodular constraint action}, when we fix $\partial_{\mu}V^{\mu}=1$, which can be achieved locally by a partial gauge fixing of diffeomorphisms. A gauge fixing prior to variation is not a generally admissible step and thus this does not guarantee the equivalence of the two theories. Nevertheless, the classical equivalence can be immediately demonstrated from the equations of motion. The variation with respect to $g_{\mu\nu}$ yields
\begin{equation}
    G_{\mu\nu}+\lambda\,g_{\mu\nu}=T_{\mu\nu}\ ,\label{sec2.2: HT einstein eom}
\end{equation}
and by varying $V^{\mu}$ we obtain
\begin{equation}
    \partial_{\mu}\lambda = 0\ .
\end{equation}
These equations are clearly the same as \eqref{sec2.1: unimodular constraint eom} and \eqref{sec2.1: constancy of cosmological constant}. The difference is that the second equation now arose as an equation of motion rather then due to Bianchi identity. Finally, the constraint associated with $\lambda$ forces
\begin{equation}
    \partial_{\mu}V^{\mu}=\sqrt{-g}\ .\label{sec2.2: HT constraint}
\end{equation}
The status of the cosmological constant problem in this formulation is very similar to the version discussed in \cref{section: unimodular constraint}. We can see that the Lagrange multiplier $\lambda$ enters the Einstein equation in the same way as in \eqref{sec2.1: unimodular constraint eom}. It is thus not surprising that determining the cosmological constant by fixing $\lambda$ directly through an initial condition will lead to the CC problem. Similarly, trying to use the constraint within the action to decouple any terms of the form
\begin{equation}
    \rho_{vac}\sqrt{-g}\ ,
\end{equation}
will yield shifts of the initial value set up for $\lambda$ exactly like in \eqref{sec2.1: initial condition shift}. On the other hand, the same solution that worked in \cref{section: unimodular constraint} works here as well. If we give up the initial condition on $\lambda$ we may find a form of the action where $\lambda$ can be determined uniquely. The steps are similar as well. We use the constraint \eqref{sec2.2: HT constraint} in the action to substitute
\begin{equation}
    g_{\mu\nu}\rightarrow \hat{g}_{\mu\nu}=g_{\mu\nu}\,\left(\frac{\partial_{\mu}V^{\mu}}{\sqrt{-g}}\right)^{1/2}\ ,\label{sec2.2: hat metric redefinition}
\end{equation}
which yields the action
\begin{equation}
    S[g,\lambda,V,\Psi]=\int_{\mathcal{M}} d^{4}x\left [\ -\frac{1}{2}\hat{R}\partial_{\mu}V^{\mu}-\lambda\left (\partial_{\mu}V^{\mu}-\sqrt{-g}\right )\right ]+S_{matter}[\hat{g},\Psi]\ .\label{sec2.2: HT action renormalized decoupled}
\end{equation}
Upon variation, this gives the following tensor equation of motion
\begin{equation}
    \hat{G}_{\mu\nu}-\frac{1}{4}\hat{G}\hat{g}_{\mu\nu}+\lambda g_{\mu\nu}=\hat{T}_{\mu\nu}-\frac{1}{4}\hat{T}\hat{g}_{\mu\nu}\ .\label{sec2.2: traceless einstein equation with cc}
\end{equation}
Taking the trace of this equation immediately gives us $\lambda=0$ and thus $\lambda$ is determined uniquely. Crucially, any quantum correction to the vacuum energy from the matter sector or the gravitational sector couples directly to $\sqrt{-\hat{g}}$, which immediately reduces it to a boundary term since the metric $\hat{g}_{\mu\nu}$ by construction satisfies
\begin{equation}
    \sqrt{-\hat{g}}=\partial_{\mu}V^{\mu}\ .\label{sec2.2: mimetic HT constraint}
\end{equation}
Hence such terms decouple trivially, which renders the cosmological constant stable under quantum corrections of the vacuum energy. Note that unlike \eqref{sec2.1: g hat definition} the metric $\hat{g}_{\mu\nu}$ is a metric in a true sense - a rank 2 tensor with a density weight of 0. Furthermore, the constraint \eqref{sec2.2: HT constraint} reduces it to
\begin{equation}
    g_{\mu\nu}=\hat{g}_{\mu\nu}\ .
\end{equation}
so we can drop the hats in our tensor equations of motion.

\subsection{Diffeomorphism covariant, Weyl invariant UG}\label{section: Weyl invariant}
Finally, similarly to \eqref{sec2.1: unimodular constraint action renormalized decoupled} the constraint in the action \eqref{sec2.2: HT action renormalized decoupled} can be omitted to give
\begin{equation}
    S[g,V,\Psi]=S_{GR}[\hat{g}(g,V),\Psi]=-\frac{1}{2}\int_{\mathcal{M}} d^{4}x \hat{R}\partial_{\mu}V^{\mu}+S_{matter}[\hat{g},\Psi]\ .\label{sec2.2: HT action renormalized decoupled no constraint}
\end{equation}
This theory amounts to ordinary GR, whose metric is transformed using the \emph{manisfestly Weyl invariant} redefinition \eqref{sec2.2: hat metric redefinition}. The Weyl symmetry of the ansatz is then inherited by the entire action \eqref{sec2.2: HT action renormalized decoupled no constraint} and thus we obtain a Weyl invariant, fully covariant theory of UG. This theory has been first suggested in \cite{Kimpton:2012rv} and was later found as a generalization of mimetic gravity in \cite{Jirousek:2018ago}, where the classical equivalence to HT formulation of UG \eqref{sec2.2: unimodular HT action} has been pointed out.

In comparison to \eqref{sec2.1: unimodular constraint action renormalized decoupled no constraint} the presence of the derivative of the vector field $V^{\mu}$ in the redefinition \eqref{sec2.2: hat metric redefinition} implies that \eqref{sec2.2: HT action renormalized decoupled no constraint} is a higher derivative vector-tensor theory. This can be easily seen by expanding the scalar curvature in the gravitational part of \eqref{sec2.2: HT action renormalized decoupled no constraint} to obtain
\begin{equation}
    S_{grav}=-\frac{1}{2}\int d^{4}x\,\sqrt{-g}\left [\sqrt{D}R+\frac{3}{8}\frac{g^{\mu\nu}\partial_{\mu}D\partial_{\nu}D}{D^{3/2}}\right]\ ,
\end{equation}
where we have denoted
\begin{equation}
    D=\frac{\partial_{\mu}V^{\mu}}{\sqrt{-g}}\ .\label{sec2.3: D factor}
\end{equation}
We can directly see that the action contains up to second time-derivatives of the time component of $V^{\mu}$. Nevertheless, due to the structure, in which these terms enter the action, and crucially, due to the universal coupling of $V^{\mu}$ to the matter sector, the equations of motion for $V^{\mu}$ reduce to a very simple form when written using $\hat{g}_{\mu\nu}$. Specifically, they become
\begin{equation}
    \partial_{\mu}\left (\hat{G}-\hat{T}\right )=0\ .\label{sec2.3: V eom}
\end{equation}
The dynamics of the vector field $V^{\mu}$, as perceived in the space-time, that is given by $\hat{g}_{\mu\nu}$, is determined by the built-in constraint \eqref{sec2.2: mimetic HT constraint}. Finally, the tensor equations of motion for $g_{\mu\nu}$ are the traceless equations \eqref{sec2: traceless einstein equation 2} evaluated using the metric $\hat{g}_{\mu\nu}$\footnote{Note that the system equations is understood as a system for the field $\hat{g}_{\mu\nu}$ rather then for $g_{\mu\nu}$}. Hence, despite the higher derivative structure of this theory, the classical dynamics are equivalent to UG \eqref{sec2.2: unimodular HT action} and the theory does not suffer from any ghost instability\footnote{Interestingly, the resulting Hamiltonian is linear in the momentum $\lambda$ and thus it is unbounded from bellow. Nevertheless, since $\lambda$ becomes a constant on-shell, the system is perfectly stable.\cite{Kluson:2014esa}}. Crucially, in contrast to \eqref{sec2.1: unimodular constraint action renormalized decoupled no constraint} the reliance of this mechanism on the extra vector field $V^{\mu}$ allows us to propose deviations from the basic theory, without affecting the decoupling mechanism for vacuum energies. For example, we can introduce novel terms in the action, which alter the dynamics of the vector field. To preserve the decoupling mechanism these terms must depend strictly on $\hat{g}_{\mu\nu}$ and on the composite vector field
\begin{equation}
    W^{\mu}=\frac{V^{\mu}}{\partial_{\nu}V^{\nu}}\ .
\end{equation}
By adding such terms the theory no longer describes UG as the original cosmological constant can acquire non-trivial dynamics and thus could become a more complicated model of dark energy, which not only models late time acceleration of the universe but also solves the old cosmological constant problem. 

The form of the redefinition \eqref{sec2.2: hat metric redefinition} is not unique in its ability to facilitate the decoupling of vacuum energies from gravity. An alternative ansatz has been proposed in \cite{Hammer:2020dqp}, where the metric $\hat{g}_{\mu\nu}$ is given through the following relation
\begin{equation}
    g_{\mu\nu}\rightarrow\hat{g}_{\mu\nu}=\frac{g_{\mu\nu}}{\sqrt[4]{-g}}\sqrt{\mathcal{P}}\label{sec2.3: axionic ansatz}
\end{equation}
where $\mathcal{P}$ is the Pontryagin density

\begin{equation}
        \mathcal{P}=\mathrm{Tr}\,\tilde{F}^{\mu\nu}F_{\mu\nu}\ ,
\end{equation}
constructed from the Yang-Mills gauge field strength
\begin{equation}
    F_{\mu\nu}=D_{\mu}A_{\nu}-D_{\nu}A_{\mu}\ .
\end{equation}
The derivative $D_{\mu}$ is the covariant derivative associated with $A_{\mu}$. Finally, $\Tilde{F}^{\mu\nu}$ is the density dual of $F_{\mu\nu}$
\begin{equation}
    \tilde{F}^{\mu\nu}=\frac{1}{2}\epsilon^{\mu\nu\rho\sigma}\,F_{\rho\sigma}\ ,
\end{equation}
and $\epsilon^{\mu\nu\rho\sigma}$ is the Levi-Civita symbol. Crucially, the mechanism by which the vacuum energies decouple is intact as the metric determinant is again constrained to\footnote{Note that in the Abelian case $\mathcal{P}$ represents the Pfaffian of the matrix $F_{\mu\nu}$}
\begin{equation}
    \sqrt{-\hat{g}}=\mathcal{P}\ ,\label{sec2.3: axionic mimetic constraint}
\end{equation}
which is a total derivative of the Chern-Simons current density $C^{\mu}$
\begin{equation}
    \mathcal{P}=\partial_{\mu}C^{\mu}\ ,\qquad\mathrm{where}\qquad C^{\mu}=\epsilon^{\mu\nu\rho\sigma}\,\mathrm{Tr}\,\left (F_{\nu\rho}A_{\sigma}-\frac{2if}{3}A_{\nu}A_{\rho}A_{\sigma}\right )\ .
\end{equation}
Here $f$ is the coupling constant of the associated gauge theory. Hence any corrections to vacuum energy are by construction decoupled as they equate to a total derivative in the Lagrangian
\begin{equation}
    \int_{\mathcal{M}} d^{4}x\,\rho_{vac}\sqrt{-\hat{g}}=\int_{\mathcal{M}} d^{4}x\,\rho_{vac}\mathcal{P}\ .\label{sec2.3: theta term}
\end{equation}
Applying \eqref{sec2.3: axionic ansatz} to GR results in the theory defined by the simple substitution as
\begin{equation}
    S[g,A,\Psi]=S_{GR}[\hat{g}(g,A),\Psi]=-\frac{1}{2}\int_{\mathcal{M}} d^{4}x \hat{R}\mathcal{P}+S_{matter}[\hat{g},\Psi]\ .\label{sec2.3: axionic substitution}
\end{equation}
Due to the Weyl invariance of \eqref{sec2.3: axionic ansatz} the tensor equations of motion are again the traceless Einstein equations \eqref{sec2: traceless einstein equation 2} evaluated using the metric $\hat{g}_{\mu\nu}$. Furthermore, the equation of motion for $A_{\mu}$ gives
\begin{equation}
    \Tilde{F}^{\mu\nu}\partial_{\nu}\left (\hat{G}-\hat{T}\right )=0\ .\label{sec2.3: A eom}
\end{equation}
For the U(1) case, the non-vanishing value of $\mathcal{P}$ implies that $\Tilde{F}^{\mu\nu}$ is an invertible matrix so we still recover \eqref{sec2.3: V eom}. For SU(N), $\Tilde{F}^{\mu\nu}$ is a Lie algebra valued object, but we find that at least one of its components (Lie algebra component) is again an invertible matrix and thus we arrive at an equivalent conclusion \eqref{sec2.3: V eom}. Hence, the gravitational dynamics are clearly equivalent to the HT formulation of UG. The dynamics of $A_{\mu}$ are determined by the built-in constraint \eqref{sec2.3: axionic mimetic constraint}. Note that solutions for this equation exist for arbitrary globally hyperbolic space-time as long as the gauge group contains an SU(2) subgroup \cite{Jirousek:2022vdq}. For the U(1) case a general proof of existence has not been found yet, but for the standard FRW space-times a solution was constructed explicitly.

There are multiple advantages of using \eqref{sec2.3: axionic ansatz} in contrast to \eqref{sec2.2: hat metric redefinition} and other formulations of UG. The gauge fields $A_{\mu}$ are clearly more natural objects in the context of the Standard Model of particle physics. Hence, the present formulation is advantageous for the study of modifications via possible couplings to ordinary matter. Such extensions could be very interesting to explore as they could provide additional dynamics that might allow us to study selection mechanisms for the otherwise unspecified effective CC in UG. Furthermore, while the quantum corrections to vacuum energy get automatically converted into a total derivative, the resulting boundary term is not necessarily physically irrelevant. In this particular case, it plays the role of the theta term \cite{CALLAN1976334,PhysRevLett.37.172} \eqref{sec2.3: theta term} for the corresponding Yang-Mills theory, under the assumption that we also include appropriate kinetic term. This then allows us to naively connect the old cosmological constant problem with the strong CP problem of quantum chromodynamics. From equation \eqref{sec2.3: A eom} it is clear that upon the addition of a kinetic term for $A_{\mu}$, the unimodular dynamics are altered, and the theory no longer describes GR with a cosmological constant \cite{Mirzagholi:2020ote}. Nevertheless, the decoupling mechanism for corrections to vacuum energy \eqref{sec2.3: theta term} is still applicable.

Finally, the theory can be written using a Lagrange multiplier in a form analogous to \eqref{sec2.2: unimodular HT action}. That is
\begin{equation}
    S[g,\lambda,A]=\int_{\mathcal{M}} d^{4}x\Big [-\frac{1}{2}\sqrt{-g}R+\lambda\big ( \mathrm{Tr}\,\tilde{F}^{\mu\nu}F_{\mu\nu}-\sqrt{-g}\big )\Big ]+S_{matter}[g,\Psi]\ .\label{sec2.3: axionic HT action}
\end{equation}
We can see that in this form the Lagrange multiplier couples to the gauge fields like an axion field. Since $\lambda$ is constrained to be a constant through the $A_{\mu}$ equation of motion we can even add a kinetic term $\partial_{\mu}\lambda\partial^{\mu}\lambda$ to the action to increase this resemblance further\footnote{Note that this preserves the original solutions while also adding a novel branch, where $\lambda$ becomes dynamical. This addition can be also applied to \eqref{sec2.2: unimodular HT action}, where the constraint on constancy of $\lambda$ is stricter and no new branch appears.}. We get
\begin{equation}
    S[g,\lambda,A]=\int_{\mathcal{M}} d^{4}x\Big [-\frac{1}{2}\sqrt{-g}R+\frac{1}{2}\sqrt{-g}g^{\mu\nu}\partial_{\mu}\lambda\partial_{\nu}\lambda+\lambda\big ( \mathrm{Tr}\,\tilde{F}^{\mu\nu}F_{\mu\nu}-\sqrt{-g}\big )\Big ]+S_{matter}[g,\Psi]\ .\label{sec2.3: axionic HT action with kinetic term}
\end{equation}
Note that this action cannot be straightforwardly traced back to a formulation without Lagrange multiplier. Nevertheless, since the kinetic term is shift symmetric, we are still able to decouple any \emph{constant} corrections to vacuum energy as long as we let the initial conditions on zero mode of $\lambda$ be free. Again adding a kinetic term for $A_{\mu}$ changes the dynamics significantly; however, it still does no affect the decoupling mechanism. It is then a attractive speculation whether unimodular gravity can arise as a dynamical regime of axion where the dynamics of the gauge field become frozen.

\subsection{Degrees of freedom of UG}\label{section: degrees of freedom}
The field content in the Henneaux and Teitelboim theory is clearly larger then in GR; however, it is accompanied by a large symmetry group of divergenceless shifts of the vector field
\begin{equation}
    V^{\mu}\rightarrow V^{\mu}+\zeta^{\mu}\ ,\qquad\mathrm{where}\qquad \partial_{\mu}\zeta^{\mu}=0\ .\label{sec3: HT gauge freedom}
\end{equation}
Consequently, much of the field content is a pure gauge and the theory can be shown to only contain a single extra global degree of freedom in comparison with its GR counterpart \cite{Henneaux:1989zc,Kluson:2014esa}. The global degree of freedom here is given as the overall charge associated with the current $V^{\mu}$, which is defined on a given foliation $\Sigma_{t}$.
\begin{equation}
    \mathcal{T}=\int_{\Sigma_{t}}d\Sigma_{\mu}V^{\mu}\ .\label{sec3: cosmic time definition}
\end{equation}
Here $d\Sigma_{\mu}=n_{\mu}\,d^{3}y$, where $d^{3}y$ is the associated coordinate volume element on $\Sigma_{t}$ and $n_{\mu}$ is perpendicular to $\Sigma_{t}$\footnote{Note that if $\Sigma_{t}$ is infinite the definition of $\mathcal{T}$ might result in an infinity as well. In such cases one will have to consider some appropriate regularization in order to make sense of these quantities.}.
$\mathcal{T}$ is often referred to as the 'cosmic time'. Note that the symmetry \eqref{sec3: HT gauge freedom} shifts $\mathcal{T}$ by a constant value
\begin{equation}
    \mathcal{T}\rightarrow \mathcal{T}+const.
\end{equation}
as long as $\zeta^{\mu}$ vanishes at infinity or on the boundary of $\Sigma_{t}$. Such shift correspond to a symmetry of the theory, which naturally results in a conservation law, in this case, the conservation of the conjugate momentum $\lambda$. Transformations that leave $\mathcal{T}$ intact form a gauge group of the theory. It follows that $\mathcal{T}$ is the only gauge-invariant information contained within the vector field $V^{\mu}$ and the gauge symmetry can be taken to fix the non-zero mode of $n_{\mu}V^{\mu}$ arbitrarily.

Integrating the expression \eqref{sec2.2: HT constraint}, while assuming appropriate conditions for $V^{\mu}$ on the boundary of $\Sigma_{t}$, enables us to calculate the change in $\mathcal{T}$ as
\begin{equation}
    \mathcal{T}(t_{2})-\mathcal{T}(t_{1})=\mathrm{Vol}_{\mathcal{M}}[g]\ ,\label{sec3: change in cosmic time}
\end{equation}
where $\partial \mathcal{M}=\Sigma_{t_{2}}\cup \Sigma_{t_{1}}$. The two variables $\lambda$ and $\mathcal{T}$ are conjugate of each other in the sense that their Dirac bracket is
\begin{equation}
    \{\lambda,\mathcal{T}\,\}_{D}=1\ .\label{sec3: commutation relation}
\end{equation}
Since the constraint part of the action \eqref{sec2.2: unimodular HT action} is linear in time derivatives we can glance this commutations directly from the action using the Fadeev-Jackiw procedure \cite{PhysRevLett.60.1692,Jackiw:1993in}. Considering only the spatially constant part of $\lambda$ we find that the only term containing time derivatives has the form
\begin{equation}
    \int d^{4}x\lambda\partial_{\mu}V^{\mu}\approx\int\,dt\lambda\,\frac{d}{dt}\int_{\Sigma_{t}}d\Sigma_{\mu}V^{\mu}=\int dt\lambda\dot{\mathcal{T}}\ .
\end{equation}
Hence we see that the momentum associated with $\mathcal{T}$ is indeed $\lambda$ and \eqref{sec3: commutation relation} immediately follows. This can be also confirmed directly using Dirac analysis. Note, that while this is true in the original HT formulation \eqref{sec2.2: unimodular HT action}, in classically equivalent actions \eqref{sec2.2: HT action renormalized decoupled} and \eqref{sec2.2: HT action renormalized decoupled no constraint} this no longer holds as $\lambda$ is constrained to vanish in the former and is not present in the latter.

In order to obtain a unique evolution in UG one has to provide both the initial cosmic time $\mathcal{T}_{i}$ as well as the value for the cosmological constant $\lambda_{i}$. In practise, $\mathcal{T}_{i}$ can be easily omitted as the evolution of the cosmic time does not affect the dynamics of gravity and other fields. It is often considered unphysical \cite{Fiol:2008vk,Padilla:2014yea}. However, we would like to point out that, while $\mathcal{T}$ is indeed unphysical, the difference of the cosmic time between two hypersurfaces $\Sigma_{t_{1}}$ and $\Sigma_{t_{2}}$ is a physical quantity, namely the total volume \cite{Percacci:2017fsy,deBrito:2021pmw}. This encodes the information about the effective cosmological constant, which can be reconstructed from the knowledge of such difference. This can be seen by fixing $\mathcal{T}(t_{2})=\mathcal{T}_{2}$ and $\mathcal{T}(t_{1})=\mathcal{T}_{1}$ in \eqref{sec3: change in cosmic time}, which yields a global constraint on the four-volume. We can then determine the value of the effective cosmological constant by solving the Einstein equation with an unspecified cosmological constant, for example the traceless equations \eqref{sec2: traceless einstein equation}, and label its solutions by the value of the cosmological constant. Hence we get a one parameter family of solutions labeled by $\Lambda$
\begin{equation}
    g_{\mu\nu}(\Lambda)\ .
\end{equation}
Plugging such solution into \eqref{sec3: change in cosmic time} with fixed values of the cosmic time yields a single equation that is in general able to determine $\Lambda$ as a function of $\Delta\mathcal{T}=\mathcal{T}_{2}-\mathcal{T}_{1}$. Let us demonstrate this explicitly on a very simple example. We consider a flat FRW universe, which is void of matter and energy, up to the unspecified cosmological constant. Hence the solutions of Friedmann equations are
\begin{equation}
    a(t)=a_{0}e^{\sqrt{\Lambda/3}(t-t_{1})}\ .\label{sec3: FRW solution}
\end{equation}
Plugging this into \eqref{sec3: change in cosmic time} yields the following relation
\begin{equation}
    \Delta\mathcal{T}=\frac{a_{0}^{3}\sqrt{3}}{\sqrt{\Lambda}}\left (e^{\sqrt{\Lambda/3}(t_{2}-t_{1})}-1\right )\ ,
\end{equation}
where we have rescaled the values of the cosmic time to factor out the infinite coordinate volume $\mathrm{Vol}_{3}=\int d^{3}x$. The above equation is an algebraic equation for $\Lambda$\footnote{This solution can be found explicitly using the Lambert $W$ function as
\begin{equation}
    \Lambda=3\,W_{0}^{2}\left (-a_{0}^{3}\frac{\Delta t}{\Delta\mathcal{T}}\right )\Delta t^{-2}\ ,\label{sec3: lambda solution}
\end{equation}
where $\Delta t= t_{2}-t_{1}$.
}. Hence the solution for the scale factor can be written as
\begin{equation}
    a(t)=a_{0}e^{\sqrt{\Lambda(\Delta\mathcal{T})/3}(t-t_{i})}\ .\label{sec3: solution based on time}
\end{equation}
Note that any correction to the value $\Lambda\rightarrow \Lambda+\rho_{vac}$, is irrelevant here. We will reproduce the above solution \eqref{sec3: solution based on time} for any value of $\rho_{vac}$ we account for. Hence specifying the effective cosmological constant by providing the value $\Delta\Lambda$ for two given times $t_{f}$ and $t_{i}$ is stable under quantum corrections. As it has been noted in \cite{Kaloper:2014dqa}, in this case "it is the space-time volume that remains fixed, forcing $\Lambda$ to adjust". Note that the global constraint \eqref{sec3: change in cosmic time} is qualitatively equivalent to the diffeomorphism invariant constraint \eqref{sec2.1: global constraint four volume}, with the difference that instead of $\mathcal{T}_{1,2}$ we are given a coordinate volume of the space-time region $\mathcal{M}$. The constraint \eqref{sec2.1: global constraint four volume} is automatically present in formulations, which do not rely on the use of the Lagrange multiplier. This implies that resolution of the CC problem in UG inherently involves the existence of such global constraint.

Note that these global constraints do not need to entail 'knowledge of the future' as both hypersurfaces $\Sigma_{t_{1},t_{2}}$ can be located in the past. This still fixes the cosmological constant, which can then be taken to determine the solutions for arbitrary future times. Note, that the hypersurfaces must not be too close to each other as the infinitesimal change in $\mathcal{T}$ becomes insensitive to the cosmological constant. Indeed, in the current setting taking the limit $t_{2}\rightarrow t_{1}$ would give us
\begin{equation}
    \dot{\mathcal{T}}=\int_{\Sigma_{t}}\,d^{3}y\sqrt{-g}\ . 
\end{equation}
which for the FRW solution \eqref{sec3: FRW solution} gives us
\begin{equation}
    \dot{\mathcal{T}}_{i}\propto a^{3}_{i}=a^{3}_{0}\ .
\end{equation}
which does not depend on $\Lambda$.

\section{Quantum aspects of UG}\label{section: quantum aspects}

As we have seen in the previous section the way we specify the initial conditions for our variables substantially affects the behavior of the theory with respect to the quantum correction of the effective cosmological constant. We have demonstrated this behavior on a semi-classical level, where the quantum corrections of the vacuum energy have been accounted only as unspecified shifts of the energy momentum tensor \eqref{sec2: vacuum shift}. The entire gravitational sector has been considered only on a classical level. In order to resolve the old CC problem to full satisfaction, one should address it in a quantum setting. Since the structure of UG is so similar to GR the problem of finding its fully quantum formulation is as problematic as that in Einstein's theory. Hence the full quantum treatment is not within our technical reach. However, the extra degrees of freedom that are present in UG have a very simple structure and can be quantized separately from the degrees of freedom of the metric and matter.

In this section we review a procedure where these degrees of freedom are integrated out in the path integral sense and by doing so they introduce a minor or no modification of the ordinary GR dynamics \cite{Fiol:2008vk,Smolin:2009ti,Padilla:2014yea,Bufalo:2015wda,Buchmuller:2022msj}. The distinction hinges on the way we carry out such integration. In particular we will explore two ways. Either we fix the initial condition for $\lambda$, or we keep $\lambda$ free. The effect of such procedure is in line with our argument from \cref{section: classical}. That is: the former way reconstructs GR, while the latter offers a resolution to the old CC problem. Finally, it has been observed that the canonical structure of the unimodular degrees of freedom implies a non-trivial commutation relations for the cosmological constant and the space-time volume. This naively implies presence of quantum fluctuations of these quantities, which could present a possible distinction between GR and UG on a quantum level.

\subsection{Path integral}\label{section: path integral}

The full formal expression for the generating function in UG for the action \eqref{sec2.2: unimodular HT action} can be written \emph{formally} as a path integral
\begin{equation}
    Z[J]=\int [Dg][D\Psi][D\lambda][DV]\exp\left (iS[g,\Psi,\lambda,V]+iS_{ext}[g,\Psi,J]\right )\ .\label{sec4: generating function}
\end{equation}
Note that we couple the external current only to the metric and matter degrees of freedom and not the fields $\lambda$ or $V^{\mu}$. The extra degrees of freedom of unimodular gravity are not deeply intertwined with the rest of the gravitational dynamic as they are neatly isolated within the constraint part of the action. Hence we can integrate them out separately prior to integration of the metric degrees of freedom or matter degrees of freedom\footnote{Note that correctly one should go first to the ADM formalism to workout the canonical structure and then calculate the associated path integral in the Hamiltonian formalism along with any necessary fixing of gauge symmetries and associated Faddeev-Popov determinants. The procedure has been carried out in ADM formalism both in \cite{Buchmuller:2022msj}, while the ghost sector has been discussed in \cite{Bufalo:2015wda}. Nevertheless, such considerations do not meaningfully affect the result in comparison to a more naive approach we consider here.}. We may thus define this partial integration as
\begin{equation}
    \mathcal{I}\equiv\int[D\lambda][DV]\exp\left(iS[g,\Psi,\lambda,V]\right )\ .\label{sec4: basic path integral}
\end{equation}
The generating functional \eqref{sec4: generating function} can then be calculated by integrating $\mathcal{I}$ over the metric and matter degrees of freedom along with external sources. The integration \eqref{sec4: basic path integral} can be carried out in more then one way depending on how we fix the initial and final condition for our fields $\lambda$ and $V^{\mu}$ or rather for the associated degree of freedom $\lambda$ and $\mathcal{T}$. We are going to be mainly looking at two ways: First, we fix the initial and final value of the cosmic time $\mathcal{T}$ and second we fix the initial and final value of $\lambda$. Note that the latter case has been worked out in \cite{Padilla:2014yea}. Technically, it is possible to fix both as it has been done in \cite{Buchmuller:2022msj}; however, we would like to point out that the knowledge of both value of $\lambda$ and $\mathcal{T}$ is prohibited due to the commutation relation \eqref{sec3: commutation relation}. Hence, the physically relevant calculation fixes only one of these variables on a given spatial slice.

We first consider the following path integral where the endpoint values for $\mathcal{T}$ are fixed
\begin{equation}
    \mathcal{I}_{\mathcal{T}}\equiv\int_{\mathcal{T}_{i}}^{\mathcal{T}_{f}}[D\lambda][DV]\exp\left(iS[g,\Psi,\lambda,V]\right )\ .\label{sec4: HT path integral cosmic time}
\end{equation}
The action in the exponent is taken to be the HT action for UG \eqref{sec2.2: unimodular HT action}. The integration over the field $V^{\mu}$ in \eqref{sec4: HT path integral cosmic time} is taken only over configurations that satisfy the following conditions

\begin{equation}
    \int_{\Sigma_{i}}d\Sigma_{\mu}V^{\mu}=T_{i}\ ,\qquad\qquad\mathrm{and}\qquad\qquad\int_{\Sigma_{f}}d\Sigma_{\mu}V^{\mu}=T_{f}\ .
\end{equation}
Note that the cosmic time $\mathcal{T}$ is the only gauge invariant information in $V^{\mu}$ and thus when we consider appropriate fixing of the symmetry \eqref{sec3: HT gauge freedom} into account the above conditions determine the initial and final configurations of $V^{\mu}$ completely. The integration over $\lambda$ is carried out freely without fixing the endpoints. As a first step we divide the action into the GR component and the constraint
\begin{equation}
    S_{HT}[g,\Psi,\lambda,V]=S_{EH}[g,\Psi]+\int_{\mathcal{M}}d^{4}x\,\lambda\left (\partial_{\mu}V^{\mu}-\sqrt{-g}\right )\ .
\end{equation}
Here $S_{EH}$ corresponds to the Einstein-Hilbert action together with arbitrary matter action for $\Psi$ in the theory. This part of the action is unaffected by the integration and thus we may focus on the constraint itself. In order to isolate the initial and final condition on the cosmic time we first integrate by parts to obtain
\begin{equation}
    S_{HT}[g,\Psi,\lambda,V]=S_{EH}[g,\Psi]+\int_{\mathcal{M}}d^{4}x\,\left (-V^{\mu}\partial_{\mu}\lambda -\lambda\sqrt{-g}\right )+\int_{\partial\mathcal{M}}d\Sigma_{\mu}\lambda\,V^{\mu}\ .\label{sec4: integrated by parts}
\end{equation}
Since the last term is evaluated on the boundary $\partial\mathcal{M}=\Sigma_{f}\cup\Sigma_{i}$ where $V^{\mu}$ is fixed, this term is unaffected by the integration over $V^{\mu}$. Hence the integration over $V^{\mu}$ gives us a delta function
\begin{equation}
   I_{\mathcal{T}}=\int_{\mathcal{T}_{i}}^{\mathcal{T}_{f}}[D\lambda]\delta(\partial_{\mu}\lambda)\exp\left(iS_{EH}[g,\Psi]-i\int_{\mathcal{M}} d^{4}x\,\sqrt{-g}\lambda+i\int_{\partial\mathcal{M}}d\Sigma_{\mu}\lambda\,V^{\mu}\right )\ .\label{sec4: delta function for lambda}
\end{equation}
The integration over the delta function fixes $\lambda$ to be a constant and thus it can be taken in front of the integral in the action. This allows us to express $V^{\mu}$ completely as the cosmic time $\mathcal{T}$
\begin{equation}
    I_{\mathcal{T}}=\int_{-\infty}^{\infty}d\lambda\exp\left(iS_{EH}[g,\Psi]-i\lambda \left(\mathcal{T}_{f}-\mathcal{T}_{i}-\int_{\mathcal{M}}d^{4}x\sqrt{-g}\right)\right )\ .\label{sec4: integration over constant lambda}
\end{equation}
Note that the delta function did not fix $\lambda$ completely and thus we are meant to integrate over the residual constant part. This gives us an ordinary delta function fixing a global constraint
\begin{equation}
    \mathcal{I}_{\mathcal{T}}=\delta(\mathcal{T}_{f}-\mathcal{T}_{i}-\mathrm{Vol}_{\mathcal{M}}[g])\exp\left(iS_{EH}[g,\Psi]\right )\ .\label{sec4: HT path integral cosmic time final}
\end{equation}
We can see that the integration over the Lagrange multiplier $\lambda$ introduces an extra global constraint on the metric volume of the considered spacetime region $\mathcal{M}$
\begin{equation}
    \mathrm{Vol}_{\mathcal{M}}[g]=\mathcal{T}_{f}-\mathcal{T}_{i}\ .
\end{equation}
From \eqref{sec4: integration over constant lambda} we can see that the Einstein Hilbert action thus obtains an unspecified cosmological constant term, which is, however, classically fixed by the global volume as we have explained in \cref{section: degrees of freedom}. Note that any shift of the vacuum energy, which we may obtain by integrating out some heavy modes of matter fields $\int[D\Psi]$ only acts to rescale the entire expression
\begin{equation}
    \mathcal{I}_{\mathcal{T}}\rightarrow \exp\left (i\rho_{vac}(\mathcal{T}_{f}-\mathcal{T}_{i})\right )\,I_{\mathcal{T}}\ ,\qquad\mathrm{as} \qquad S_{EH}[g,\Psi]\rightarrow S_{EH}[g,\Psi]+\rho_{vac}\mathrm{Vol}_{\mathcal{M}}[g]
\end{equation}
Hence correlation functions of any kind remain unaffected by such shifts and consequently local measurements are unaffected as well.

Now we consider the situation where we fix the endpoint values of $\lambda$. This gives a nearly identical expression to \eqref{sec4: HT path integral cosmic time}
\begin{equation}
    \mathcal{I}_{\lambda}=\int_{\lambda_{i}}^{\lambda_{f}}[D\lambda][DV]\exp\left(iS[g,\Psi,\lambda,V]\right )\ ;\label{sec4: HT path integral lambda}
\end{equation}
however, in order to get a consistent result we must alter the action that we use. Note that in \eqref{sec2.2: unimodular HT action} one must pose a vanishing boundary conditions for the field $V^{\mu}$. This corresponds to a fixing of initial and final configuration of $V^{\mu}$ and consequently of $\mathcal{T}$, but not $\lambda$. Such action is thus suitable to calculate path integrals with fixed initial and final $\mathcal{T}$. The appropriate action to calculate the transition amplitude between $\lambda$ eigenstates is instead \eqref{sec4: integrated by parts} with the boundary term being dropped \cite{Fiol:2008vk}. Hence
\begin{equation}
    S[g,\Psi,\lambda,V]=S_{EH}[g,\Psi]+\int_{\mathcal{M}}d^{4}x\,\left (-V^{\mu}\partial_{\mu}\lambda -\lambda\sqrt{-g}\right )\ .
\end{equation}
In this action we must instead pose the vanishing of variation of $\lambda$ in order to obtain equations of motion. The integration over $V^{\mu}$ in \eqref{sec4: HT path integral lambda} is completely free while the integration over $\lambda$ has fixed endpoints
\begin{equation}
    \lambda(t_{i})=\lambda_{i}\ ,\qquad\qquad\mathrm{and}\qquad\qquad\lambda(t_{f})=\lambda_{f}\ .
\end{equation}
We can carry out the integration over $V^{\mu}$ directly to obtain
\begin{equation}
    \mathcal{I}_{\lambda}=\int_{\lambda_{i}}^{\lambda_{f}}[D\lambda]\delta(\partial_{\mu}\lambda)\exp\left(iS_{GR}[g,\Psi]-i\lambda\mathrm{Vol}_{\mathcal{M}}[g]\right )\ .
\end{equation}
The integration over the delta function now gives
\begin{equation}
    \int_{\lambda_{i}}^{\lambda_{f}}[D\lambda]\delta(\partial_{\mu}\lambda)=\delta(\lambda_{f}-\lambda_{i})\ .
\end{equation}
Hence, we obtain
\begin{equation}
    \mathcal{I}_{\lambda}=\delta(\lambda_{f}-\lambda_{i})\exp\left(iS_{GR}[g,\Psi]-i\lambda_{i}\mathrm{Vol}_{\mathcal{M}}[g]\right )\ .\label{sec4: HT path integral lambda final}
\end{equation}
In this case the cosmological constant is specified directly by $\lambda_{i}$. Consequently, the effective cosmological constant receives any shifts of vacuum energy $\rho_{vac}$ from the matter sector.

We can see that the results \eqref{sec4: HT path integral cosmic time final} is insensitive to quantum corrections of the vacuum energy while the latter is \eqref{sec4: HT path integral lambda final}. Crucially, the difference in considerations that lead to these results is exactly in line with the semi-classical case that we have discussed in \cref{section: classical}. In particular, choosing the initial value for the Lagrange multiplier $\lambda$ spoils the solution of the old cosmological constant problem. Instead, allowing $\lambda$ to be free yields a formulation where the effective cosmological constant is stable against radiative corrections. In the semi-classical case this corresponds to solving for $\lambda$ algebraically, without initial conditions, while in the present setting it corresponds to integration over the Lagrange multiplier including its zero mode. This distinction is consistent with other results on various aspects of quantum UG. For example the works \cite{Padilla:2014yea,Fiol:2008vk,Buchmuller:2022msj,Bufalo:2015wda} fix $\lambda$ by hand and the results point toward the conclusion that the status of the cosmological constant in UG is not any different from GR. On the other hand the works \cite{Smolin:2009ti,Alvarez:2015pla,Alvarez:2015sba,Percacci:2017fsy,deLeonArdon:2017qzg,Herrero-Valea:2020xaq} base their calculations on formulations that do not rely on a Lagrange multiplier to enforce \eqref{sec2.1: unimodular constraint} and their conclusions are consistent with the old CC problem indeed being solved in UG.

\subsection{Quantum fluctuations}\label{section: fluctuations}
As we have seen in \cref{section: degrees of freedom} the two global quantities $\mathcal{T}$ and $\lambda$ form a conjugate pair, with the following Dirac bracket relation
\begin{equation}
    \{\mathcal{T},\lambda\}=1\ .
\end{equation}
Upon standard canonical quantization such relation becomes a commutator of operators due to the correspondence principle
\begin{equation}
    [\hat{\mathcal{T}},\hat{\lambda}]=1\ .
\end{equation}
Consequently, the two associated observables are not simultaneously measurable and the corresponding quantities are subjected to quantum fluctuations, whose size is constrained by the Heisenberg uncertainty relations
\begin{equation}
    \delta\mathcal{T}\times\delta\lambda\geq \frac{1}{2}\ .
\end{equation}
Since the measurement of cosmic time between two hypersurfaces corresponds to the spacetime four-volume, any uncertainty in the measurement of $\mathcal{T}$ is translated to an uncertainty of the four-volume itself. Hence we can write \cite{Jirousek:2020vhy,Vikman:2021god}
\begin{equation}
    \delta\lambda\times\delta\mathrm{Vol}_{\mathcal{M}}[g]\geq\frac{1}{2}\ .
\end{equation}
Such fluctuations are mostly harmless as the four-volume is typically very large and any fluctuations in it can be localized very far from a local observer. Potentially, even in a causally disconnected region. Hence, we may usually measure $\lambda$ with an arbitrary precision. It follows that such fluctuations are unlikely to have any effect in our Universe; however, they present a conceptual difference between quantum GR and UG.

Nevertheless, we can imagine situations, where such fluctuations can have significant effects. Consider a closed, radiation dominated Friedman universe. The associated scale factor hence evolves as
\begin{equation}
    a(\eta)=a_{m}\sin(\eta)\ ,
\end{equation}
where $a_{m}$ is the scale factor at the turning point and $\eta$ is the conformal time. It is reasonable to assume that the fluctuations are smaller then the total four volume. Hence, we obtain \cite{Vikman:2021god}
\begin{equation}
    \delta\mathcal{T}<\mathrm{Vol}_{4}[g]=\frac{3\pi^{3}}{4}a^{4}_{m}\ .
\end{equation}
Using the uncertainty relation we find a lower bound on the fluctuations of the cosmological constant
\begin{equation}
    \delta\lambda>\frac{2}{3\pi^{3}}a^{-4}_{m}\ .
\end{equation}
Clearly, this is negligible in large universe but it renders small universes inconsistent as large fluctuations of $\lambda$ violate the assumption of radiation domination. It would be interesting to see whether such a small universe, that would quickly collapse in the ordinary GR setting, could grow large due to such fluctuation in $\lambda$. It would also be of interest if such fluctuations can be recovered through the path integral techniques, which have been explored in greater detail. Finally, we would like to note that just a slight modification of the Henneaux and Teitelboim construction \eqref{sec2.2: unimodular HT action} can be used to promote any physical constant to a degree of freedom by promoting the constant $\alpha$ to a scalar field and introducing the term
\begin{equation}
    V^{\mu}\partial_{\mu}\alpha\ .
\end{equation}
By extension we can obtain Heisenberg relations for various constants with their associated global conjugates. This has been performed for the Planck mass and the Planck constant \cite{Jirousek:2020vhy} as well as various others \cite{Magueijo:2021pvq,Magueijo:2021rpi,Gielen:2022ouk}

\section{Vacuum energy sequestering}\label{section: sequestering}

Another notable theory that aims to address the old cosmological constant problem is vacuum energy sequestering \cite{Kaloper:2013zca,Kaloper:2014dqa}. This proposal shares multiple similarities with UG, in particular in its local formulation \cite{Kaloper:2015jra} and hence it is useful to compare the two here. The original idea relies on an introduction of global mechanics, which enforce a pair of global constraints. These constraints then determine the effective cosmological constant in a manner that is stable against quantum corrections of vacuum energy. The global dynamics are introduced by considering a pair of \emph{global variables} $\theta$ and $\Lambda$. The former is input by hand as a rescaling of the physical metric in the gravitational sector
\begin{equation}
    g_{\mu\nu}\rightarrow \Tilde{g}_{\mu\nu}=\theta^{-2}\,g_{\mu\nu}\ .
\end{equation}
The second variable is the cosmological constant of GR promoted to an independent variable with no space-time dependence. Hence the \emph{local} part of the action is modified as
\begin{equation}
    S=\int d^{4}x\sqrt{-g}\left [-\frac{1}{2\theta^{2}}R+\Lambda+\mathcal{L}(g_{\mu\nu},\Psi) \right]\ .
\end{equation}
Furthermore, this action is supplemented by a \emph{global} term
\begin{equation}
    \sigma\left(\frac{\Lambda}{\mu^{4}}\right)\ .
\end{equation}
where $\sigma$ is an arbitrary monotonous function and $\mu$ is an unspecified dimensionful parameter, which is meant to be measured. The total action is
\begin{equation}
    S[g,\Psi,\Lambda,\theta]=\int d^{4}x\sqrt{-g}\left [-\frac{1}{2\theta^{2}}R+\Lambda+\mathcal{L}(g_{\mu\nu},\Psi) \right]+\sigma\left(\frac{\Lambda}{\mu^{4}}\right)\ .\label{sec5: sequestering action global}
\end{equation}
Crucially, the novel variables $\theta$ and $\Lambda$ are not fields and have no space-time dependence, yet, they are subjected to the variation principle. Consequently, their equations of motion yield two global constraints
\begin{equation}
    \int d^{4}x\sqrt{-g}R=0\ ,\qquad\qquad\qquad \frac{\sigma'}{\mu^{4}}=\int d^{4}x\,\sqrt{-g}\ .\label{sec5: global constraints}
\end{equation}
It is useful to introduce a space-time average of a scalar quantity as $\braket{\phi}\equiv \int d^{4}x\sqrt{-g}\phi/\int d^{4}x\sqrt{-g}$. Using this we can rewrite the first constraint as
\begin{equation}
    \braket{R}=0\ .\label{sec5: global constraint curvature}
\end{equation}
The equations of motion for the metric $g_{\mu\nu}$ are given as
\begin{equation}
    \theta^{-2}G_{\mu\nu}=T_{\mu\nu}-\Lambda g_{\mu\nu}\ .\label{sec5.1: einstein equation jordan}
\end{equation}
We can clearly see that these are just Einstein equations with an unspecified cosmological constant and unspecified rescaling of the Planck mass. The key property of the sequestering mechanism is that the global constraints \eqref{sec5: global constraints} allow us to find an explicit expression for $\Lambda$, that does not reduce the Einstein equation to traceless equations \eqref{sec2: traceless einstein equation}. This is achieved by taking the trace and a space-time average of \eqref{sec5.1: einstein equation jordan}. Doing so we obtain
\begin{equation}
    -\braket{R}=\braket{T}-4\Lambda\ .\label{sec5: averaged Einstein equations}
\end{equation}
We can use the first of the two global constraints \eqref{sec5: global constraints} to eliminate the average curvature to find\footnote{Interestingly similar constraint for $\Lambda$ has been considered in \cite{vanderBij:1981ym,Smolin:2009ti}}
\begin{equation}
    \Lambda=\frac{1}{4}\braket{T}\ .\label{sec5: global constraint lambda 2}
\end{equation}
This can be plugged back into the Einstein equation \eqref{sec5.1: einstein equation jordan}, which now reads
\begin{equation}
    \theta^{-2}G_{\mu\nu}=T_{\mu\nu}-\frac{1}{4}\braket{T}g_{\mu\nu}\ .\label{sec5.1: einstein equation sequestered}
\end{equation}
This equation clearly possesses the same symmetry \eqref{sec2: vacuum shift} as UG. However, unlike UG the covariant divergence of this equation vanishes identically. Hence, there is no differential constraint, which would give rise to an additional component of the cosmological constant. Instead, we obtain full set of 10 equations. The form of these equations \eqref{sec5.1: einstein equation sequestered} is rather unusual as it contains a term that is non-local in time. Hence, it would seem it would be difficult to interpret it as an evolution equation for a set of initial data. Nevertheless, finding solutions of these equations is rather straightforward. The method is exactly the same as we have discussed in \cref{section: degrees of freedom}. We consider the Einstein equation with an \emph{unspecified} cosmological constant $\lambda$ and the parameter $\theta$
\begin{equation}
    \theta^{-2}G_{\mu\nu}=T_{\mu\nu}-\lambda g_{\mu\nu}\ ,
\end{equation}
and find a family of solutions labeled by their values: $g_{\mu\nu}(\theta,\lambda)$. For such solutions we evaluate the energy momentum tensor $T_{\mu\nu}(\theta,\lambda)$ and calculate its associated space-time average. Plugging such expression into \eqref{sec5: global constraint lambda 2} and setting $\Lambda=\lambda$ we find a consistency equation
\begin{equation}
    \Lambda=\frac{1}{4}\braket{T}(\theta,\Lambda)\ .\label{sec5: consistency}
\end{equation}
The actual value of $\Lambda$ is then selected as a solution of this equation in terms of $\theta$. Note that existence of a solution is not in general guaranteed. If this occurs then the entire family of solutions $g_{\mu\nu}(\theta,\Lambda)$ are not solutions of equations \eqref{sec5.1: einstein equation sequestered}. Once we find $\Lambda$ we can plug it back into the second global constraint in \eqref{sec5: global constraints} in order to determine $\theta$. Crucially, unlike UG the vacuum energy sequester does not allow for arbitrary value of the cosmological constant but a specific one which is determined by the above procedure.

\subsection{Local formulation}
The main disadvantage of the sequestering proposal is that its formulation requires an unusual global term and variables. To remedy this, a local formulation of the theory has been proposed in \cite{Kaloper:2015jra}, which can be obtained by a 'localization' of the global dynamics of \eqref{sec5: sequestering action global}. The strategy to localize action \eqref{sec5: sequestering action global} is rather simple. The global variables $\theta$ and $\Lambda$ are promoted to local variables - scalar fields
\begin{equation}
    \Lambda\rightarrow\Lambda(x)\ ,\qquad\qquad\theta\rightarrow\theta(x)\ ,
\end{equation}
however, their non-constant part is immediately constrained to vanish using vector Lagrange multipliers
\begin{equation}
    V^{\nu}\,\partial_{\nu}\sigma\left(\frac{\Lambda}{\mu^{4}}\right)\ ,\qquad\qquad W^{\nu}\,\partial_{\nu}\Tilde{\sigma}\left(\frac{1}{\theta }\right)\ ,
\end{equation}
where $\sigma$ and $\Tilde{\sigma}$ are taken to be monotonous functions and $V^{\mu}$ and $W^{\mu}$ are vector densities. Hence, the equations of motion for $V^{\mu}$ and $W^{\mu}$ yield
\begin{equation}
    \partial_{\mu}\,\Lambda=\partial_{\mu}\,\theta=0\ .
\end{equation}
$\Lambda$ and $\theta$ thus become global variables only on equations of motions rather then apriori. Including the constraint terms in the action yields
\begin{align*}
    S[g,\Psi,\Lambda,\theta]=\int d^{4}x\sqrt{-g}&\left [ -\frac{1}{2\theta^{2}}R-\Lambda+\mathcal{L}(g_{\mu\nu},\Psi) \right ]\\
    &+\int d^{4}x\left [\partial_{\mu}V^{\mu}\,\sigma\left(\frac{\Lambda}{\mu^{4}}\right)+\partial_{\mu}W^{\mu}\,\Tilde{\sigma}\left(\frac{1}{\theta }\right)\right ] .\numberthis\label{sec5: sequestering action local}
\end{align*}
The extra constraint terms are metric independent and thus they do not affect the gravitational equations; however, we get new relations that govern the dynamics of the extra fields. In total, we obtain the following set of equations
\begin{align}
    \theta^{-2}G_{\mu\nu}=&\,T_{\mu\nu}-\Lambda g_{\mu\nu}\ ,\label{sec5.2: local einstein equation sequestering}\\
    \frac{\sigma'}{\mu^{4}}\partial_{\mu}V^{\mu}=\sqrt{-g}\ ,\qquad&\qquad \frac{\tilde{\sigma}'}{\theta }\partial_{\mu}W^{\mu}=\sqrt{-g}R\ ,\label{sec5.2: vector equations}\\
    \partial_{\mu}\,\Lambda=0\ ,\qquad&\qquad\partial_{\mu}\,\theta=0\ .\label{sec5.2: lambda and theta constancy}
\end{align}
We can see that the action \eqref{sec5: sequestering action local}, as well as the associated equations of motion, bear a strong resemblance to unimodular gravity. The tensor equation of motion remains unaffected by the change in description and retains its form \eqref{sec5.1: einstein equation jordan}. The vector field $V^{\mu}$ has the same role as it had in the HT formulation of UG - to force constancy of $\Lambda$. Analogously, the vector density $W^{\mu}$ is used to force constancy of $\theta$. In essence, the effective Planck mass and the cosmological constant are promoted to an integration constant rather then bare coupling constants. The physical interpretation and pitfalls of this theory are consequently very similar to UG. Indeed, if we provide initial conditions for $\Lambda$ and $\theta$ directly in order to solve \eqref{sec5.2: lambda and theta constancy}, we obtain an ordinary Einstein equation \eqref{sec5.2: local einstein equation sequestering} with the chosen constants. Such approach is clearly no different from choosing the cosmological constant directly in UG, and hence, it will be unstable against radiative corrections.

However, similar to UG, the cosmological constant can be prescribed in a stable manner. To demonstrate this we first analyze the equations of motion for $\theta$ and $\Lambda$ \eqref{sec5.2: vector equations}. These are local analogues of the global constraints \eqref{sec5: global constraints}. While being local, these equations describe the evolution of a two global quantities. Namely the 'cosmic times' $\mathcal{T}$ and $\mathcal{\Tilde{T}}$ associated with $V^{\mu}$ and $W^{\mu}$, which can be introduced as in \eqref{sec3: cosmic time definition}. Such quantities are sourced by the space-time volume and the integrated curvature respectively
\begin{equation}
    \mathcal{T}_{t_{2}}-\mathcal{T}_{t_{1}}=\frac{\mu^{4}}{\sigma'}\mathrm{Vol}_{\mathcal{M}}[g]\ ,\qquad\qquad\tilde{\mathcal{T}}_{t_{2}}-\tilde{\mathcal{T}}_{t_{1}}=\frac{\theta }{\tilde{\sigma}'}\int_{\mathcal{M}}d^{4}x\sqrt{-g}R\ .\label{sec5.2: sequestering cosmic times}
\end{equation}
These equations are now global equations, which can be used in the same manner as the original global constraints \eqref{sec5: global constraints}. In particular, consider taking the trace and a space-time average of the equation \eqref{sec5.2: local einstein equation sequestering}. We again find \eqref{sec5: averaged Einstein equations}. By taking the ratio of \eqref{sec5.2: sequestering cosmic times} we can express the averaged curvature
\begin{equation}
    \braket{R}=\frac{\sigma'}{\Tilde{\sigma}'}\,\frac{1}{\theta\mu^{4}}\,\frac{\tilde{\mathcal{T}}_{t_{2}}-\tilde{\mathcal{T}}_{t_{1}}}{\mathcal{T}_{t_{2}}-\mathcal{T}_{t_{1}}}\ .
\end{equation}
Hence from \eqref{sec5: averaged Einstein equations} we find
\begin{equation}
    \Lambda=\frac{1}{4}\braket{T}+\Delta\Lambda\ ,\label{sec5.1: Lambda equation}
\end{equation}
where
\begin{equation}
    \Delta \Lambda=\frac{\sigma'}{\Tilde{\sigma}'}\,\frac{1}{\theta^{3}\mu^{4}}\,\frac{\tilde{\mathcal{T}}_{t_{2}}-\tilde{\mathcal{T}}_{t_{1}}}{\mathcal{T}_{t_{2}}-\mathcal{T}_{t_{1}}}\ .\label{sec5.2: delta lambda}
\end{equation}
Note that \eqref{sec5.1: Lambda equation} is not in general an explicit solution for $\Lambda$ as $\sigma$ is a function of $\Lambda$. Nevertheless, plugging this expression into the Einstein equation \eqref{sec5.2: local einstein equation sequestering} we find the sequestered equations \eqref{sec5.1: einstein equation jordan} up to an extra vacuum energy piece $\Delta\Lambda$
\begin{equation}
    \theta^{-2}G_{\mu\nu}=\,T_{\mu\nu}-\frac{1}{4}\braket{T}g_{\mu\nu}-\Delta\Lambda g_{\mu\nu}\ .\label{sec5.2: einstein equation sequestering local delta lambda}
\end{equation}
This equation again has the shift symmetry \eqref{sec2: vacuum shift}, which cancels out the quantum corrections of vacuum energy on the right hand side. Furthermore, the extra piece of the effective cosmological constant, $\Delta\Lambda$, does not depend on the energy and momentum of matter at all. It depends only on gravitational quantities such as the integral of $V^{\mu}$ and $W^{\mu}$, which are sourced by the four-volume and the scalar curvature. Thus $\Delta\Lambda$ does not directly carry the information about the energy and momentum of matter. Consequently, since gravitational degrees of freedom are protected via the symmetry \eqref{sec2: vacuum shift} in the tensor equation, $\Delta\Lambda$ does not receive any such corrections. Note that unlike in the original global version of the sequester, we do not have a direct expression for the effective cosmological constant. Instead, such constant must be determined through measurement. The main point of the above discussion is to demonstrate that such value is then stable under the radiative corrections.

We would like to comment here that the equations of the local version of the sequester can be reduced to the original \eqref{sec5.1: einstein equation sequestered} if we allow ourselves prescribe initial and final values of the cosmic times $\mathcal{T}$ and $\tilde{\mathcal{T}}$. In particular, the choice $\tilde{\mathcal{T}}_{t_{2}}=\tilde{\mathcal{T}}_{t_{1}}$ reduces the second equation \eqref{sec5.2: sequestering cosmic times} to the constraint \eqref{sec5: global constraint curvature}. Equivalently, we get $\Delta\Lambda=0$ so equations \eqref{sec5.2: einstein equation sequestering local delta lambda} reduce exactly to \eqref{sec5.1: einstein equation sequestered}. Finally, we would like to point out that the same strategy can be applied to the HT formulation of UG. Providing initial and final $\mathcal{T}$ gives us a global constraint
\begin{equation}
    \mathcal{T}_{t_{2}}-\mathcal{T}_{t_{1}}=\mathrm{Vol}_{\mathcal{M}}[g]\ ,
\end{equation}
which can be used to determine the effective cosmological constant. Such constant is then clearly insensitive to any quantum corrections. As the space-time volume is fixed.

Finally, the localization procedure described in this section can be reversed and applied to unimodular gravity \eqref{sec2.2: unimodular HT action} to find a unimodular analogue of the sequestering mechanism. Doing so implies that $\lambda$ becomes a global variable
\begin{equation}
    \lambda(x)\rightarrow \lambda\ .
\end{equation}
This allows us to integrate the divergence of the vector in the constraint part of the action \eqref{sec2.2: unimodular HT action} to obtain
\begin{equation}
    S_{const}=\lambda\left (\mathcal{T}_{f}-\mathcal{T}_{i}-\mathrm{Vol}_{\mathcal{M}}[g]\right)\ .\label{sec5: global formulation of UG}
\end{equation}
The variation of the global $\lambda$ now implies \eqref{sec3: change in cosmic time}, where the time at the endpoints must be apriori specified. Upon variation this yields Einstein equations with a cosmological constant that is determined through the global constraint \eqref{sec3: change in cosmic time}. Note that the crucial difference in comparison to sequestering is that the freedom in choosing $\mathcal{T}_{f}-\mathcal{T}_{i}$ allows us to reconstruct \emph{any} value of the cosmological constant. In sequestering the global constraint that determines $\Lambda$ \eqref{sec5: global constraint curvature} does not present any choice. On the other hand the solution for $\theta$ is affected by the choice of the function $\sigma$.

\section{Conclusions}\label{section: conclusion}
In this work we discussed whether unimodular gravity is or is not able to reconcile the old cosmological constant problem. In \cref{section: classical} we pointed out that the answer hinges on a rather minute technicality - on how one provides the data that determine the effective cosmological constant. This point is completely mute on a classical level; however, it becomes crucial on a semi-classical level, when we introduce quantum corrections to vacuum energy. The distinction arises in theories, which use a Lagrange multiplier \eqref{sec2.1: unimodular constraint action}, \eqref{sec2.2: unimodular HT action} in order to enforce their respective constraints \eqref{sec2.1: unimodular constraint}, \eqref{sec2.2: HT constraint}. In such formulations one often encounters that the initial condition is set up for the Lagrange multiplier directly. Such fixing implies that the zero mode of the multiplier is not varied in the action and hence, the associated constraint is enforced only locally. The local versions of the constraints are, however, nearly 'empty' as GR possesses enough gauge symmetry to satisfy them without any effect on the dynamics. Consequently, setting up the cosmological constant in this manner amounts to little to no change in the dynamics in comparison to GR with a chosen CC. Hence, we found that the cosmological constant problem is still present when CC is chosen in this way. Leaving the initial conditions of the Lagrange multiplier be free, implies that the constraints are enforced fully. This introduces a global constraint on the four-volume \eqref{sec2.1: global constraint four volume}, \eqref{sec3: change in cosmic time}, which is able to fix the effective cosmological constant in a manner that is stable against quantum corrections \eqref{sec3: solution based on time}. Hence, such route offers a resolution of the old cosmological constant problem. The above mentioned issues are not encountered in theories, where the metric is endowed with a composite structure, which enforces the appropriate constraints \eqref{sec2.1: unimodular constraint action renormalized decoupled}, \eqref{sec2.2: HT action renormalized decoupled no constraint} automatically. When there are no Lagrange multipliers, we cannot assign initial values to them. For this reason such formulations can be considered to have an advantage over the Lagrange multiplier ones and indeed the cosmological constant problem has been reported to be solved in these versions of UG \cite{Alvarez:2015pla,Alvarez:2015sba}. 

We discussed a recently proposed pair of theories of UG \eqref{sec2.2: HT action renormalized decoupled no constraint}, \eqref{sec2.3: axionic substitution} \cite{Jirousek:2018ago,Hammer:2020dqp} in \cref{section: Weyl invariant}. These proposals combine many desired properties as they are fully diffeomorphism covariant, Weyl invariant theories of UG that do not rely on a Lagrange multiplier. Hence the cosmological constant problem is unambiguously solved within them. We discuss possible extensions of these theories beyond unimodular gravity and how they can fit within the Standard Model of particle physics, while making sure that the decoupling mechanism for quantum corrections of vacuum energy is still functional. We point out a striking similarity of the proposal \eqref{sec2.3: axionic HT action} to the axion dynamics of SU(3) Yang-Mills theory.

In \cref{section: quantum aspects} we reviewed the path integral quantization of the unimodular degree of freedom, the cosmological constant, in the HT formulation \eqref{sec2.2: unimodular HT action} with Lagrange multiplier. The structure of the additional degree of freedom is very simple and can be integrated out easily separately from the metric and matter degrees of freedom. Such integration can be carried out in two ways: by either fixing the initial and final value of the cosmological constant itself or by doing the same for its conjugate quantity - the cosmic time \eqref{sec3: cosmic time definition}. This is a direct analogue of the initial value ambiguity in the semi-classical case discussed in \cref{section: classical} and leads to the same conclusion. That is, choosing the initial value of the cosmological constant directly, spoils the solution of the cosmological constant problem. Conversely, the second route, fixing the cosmic time, leads to a reconciliation of the CC problem by introducing a global constraint. We further discuss that the promotion of the cosmological constant to a degree of freedom naively leads to an appearance of global fluctuations of the cosmological constant. Such fluctuations have likely no effect in our Universe due to its large size. Nevertheless, the existence of these fluctuations presents a conceptual difference of UG from GR.

Finally, we discussed the vacuum energy sequestering \cite{Kaloper:2013zca} in \cref{section: sequestering}. We reviewed its basic formulation and compared its working with UG. In our view, the mechanism that allows UG to alleviate the CC problem is surprisingly similar to the mechanism of vacuum energy sequestering in that the two theories can be both viewed as operating using a global constraints. In contrast to UG, the original sequestering proposal does not allow us to stray away from this global structure and thus it is guaranteed to provide a reconciliation of the old cosmological constant problem. Furthermore, in comparison to UG, the constraint, which determines the cosmological constant in sequestering is uniquely fixed. In this sense sequestering is more constrained than UG. The similarities between sequestering and UG are even more pronounced in the local formulation \eqref{sec5: sequestering action local}, which unfortunately introduces the same ambiguity in providing the initial value for the cosmological constant. The relation between the local and global formulation of sequestering can be extrapolated to allow us to write down a formulation of UG analogous to the global vacuum energy sequestering \eqref{sec5: global formulation of UG}.

In our view unimodular gravity indeed offers a resolution of the old cosmological constant problem. However, only as long as one is careful in setting up the initial value for the cosmological constant in a correct way. This particular distinction goes beyond the classical consideration, which leads to conflicting reports on the viability of UG in regards to the old CC problem. However, these findings are consistent when the above distinction is highlighted.



\vspace{6pt} 




\funding{P.~J. acknowledges funding from the South African Research Chairs Initiative of the Department of Science and Technology and the National Research Foundation of South Africa.}

\dataavailability{Not applicable.} 

\acknowledgments{It is a pleasure to thank Alexander Vikman and Ippocratis Saltas for useful discussions.}

\conflictsofinterest{The authors declare no conflict of interest. The funders had no role in the design of the study; in the collection, analyses, or interpretation of data; in the writing of the manuscript; or in the decision to publish the~results.} 



\abbreviations{Abbreviations}{
The following abbreviations are used in this manuscript:\\

\noindent 
\begin{tabular}{@{}ll}
CC & cosmological constant\\
GR & general relativity\\
UG & unimodular gravity\\
HT & Henneaux and Teitelboim\\
EH & Einstein-Hilbert\\

\end{tabular}
}




\begin{adjustwidth}{-\extralength}{0cm}

\reftitle{References}


\bibliography{references} 

%


\PublishersNote{}
\end{adjustwidth}
\end{document}